# A LOW-LATENCY INTERACTIVE SYSTEM FOR REAL-TIME VIDEO UNDERSTANDING BASED ON VLMS

Punan Dai[1], Jun Xu[1], Bingcong Lu[1], Zhengxue Cheng[1], Hongwei Hu[2], Ronghua Wu[2], Li Song[1] *

[1]School of Information Science and Electronic Engineering, Shanghai Jiao Tong University, China and [2]Ant Group, Shanghai, China

## ABSTRACT

Vision-language models are extending video understanding from offline clip analysis to continuous interactive streaming, but most research still emphasizes model capability rather than deployable low-latency interaction. This paper presents a unified edge-cloud system for real-time video VLM applications. Lightweight phone, smart glasses, PC, and pseudo-replay clients publish video and speech to a server runtime that provides shared ASR/TTS, session orchestration, backend adaptation, response delivery, and archive-backed measurement. The system integrates six representative video VLM backends with streaming or interaction-oriented capabilities and evaluates them across backend runtime, media transport, client-observed latency, and interaction behavior. With suitable backend selection and the WebRTC path, the tested system reaches approximately 0.9–1.0 s to first VLM text and 1.3–1.5 s to first non-silent TTS audio, while exposing backend adaptation costs and differences in real-time interaction behavior.

## INTRODUCTION

Video-language models have advanced rapidly in temporal reasoning, long-form video comprehension, and multimodal evaluation (Tang et al. (1), Wu et al. (2), Fu et al. (3)). These advances make VLMs attractive for broadcast assistance, immersive media production, wearable guidance, mobile monitoring, and PC-based control surfaces. These scenarios require the system to respond while the user is still acting or the scene is still evolving, rather than after a complete clip has been uploaded and analyzed.

Most video VLM studies, however, still assume an offline or near-offline setting in which a clip is available before inference, the user asks after the visual evidence has been collected, and evaluation focuses on answer accuracy. Real-time video understanding is harder in three ways. The model must maintain useful visual context under bounded computation, the interaction loop must decide when and how to answer through speech or text, and the deployed system must control transport, endpointing, synthesis, playback, and latency measurement across clients and servers. Recent VLM studies for streaming and interaction-oriented video understanding have begun to address this shift by adding streaming state, memory compression, online token management, response timing, and simultaneous perception and generation (Xu et al. (4), Yao et al. (5), Lu et al. (6), Shi et al. (7), Lin et al. (8)). These methods improve the model side of streaming understanding, but they are usually studied as model algorithms or benchmark results. Deployment still requires a system that connects lightweight terminals, bidirectional media transport, speech input and output, heterogeneous VLM backends, and reproducible measurement.

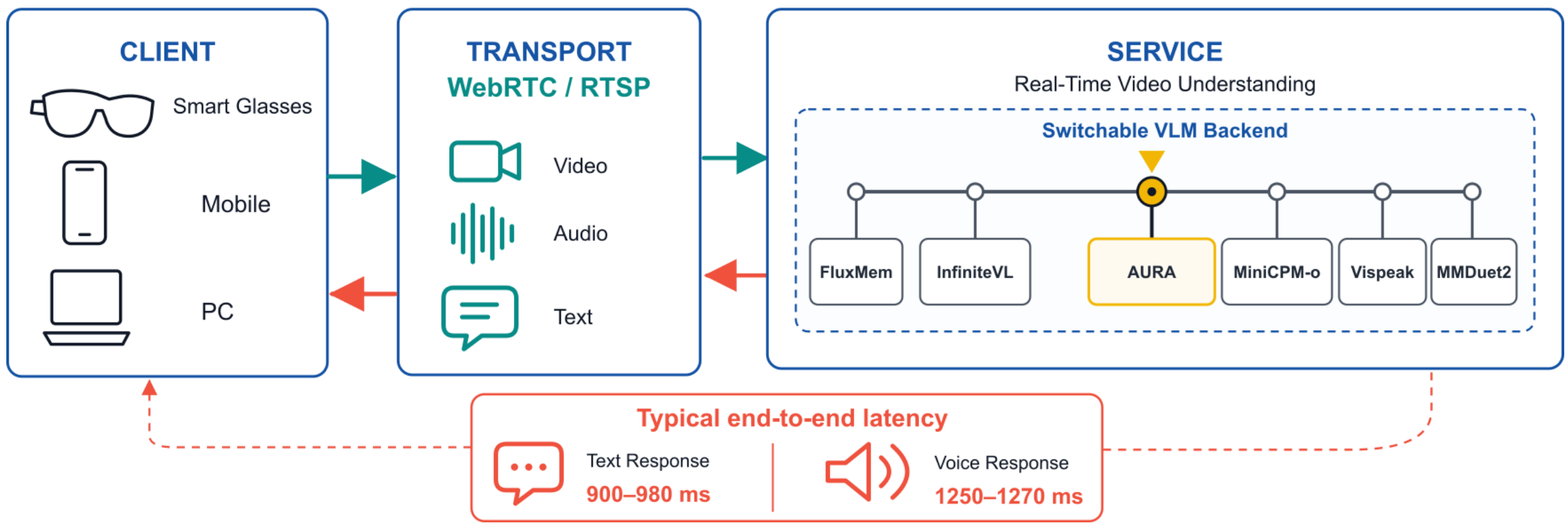

Figure 1 – System at a glance. The proposed client–transport–service architecture supports lightweight multimodal interaction, switchable integration of six VLM backends, and low-latency text and speech feedback.

This paper presents such a system. We build a low-latency edge-cloud architecture in which phone, smart glasses, PC, and pseudo-replay clients publish video and speech to a server runtime. The server provides shared ASR and TTS, session orchestration, backend adaptation, response delivery, and archive-backed measurement. The system centers on streaming video understanding but exposes a common interaction path, allowing different VLM backends to be compared under the same speech-enabled runtime.

The contributions of this paper are as follows.

1) We design and implement a low-latency edge-cloud system for real-time video VLM interaction, using lightweight clients, shared ASR/TTS services, unified session orchestration, and backend adapters.

2) We integrate six representative video VLM backends designed for streaming video understanding in the same speech-enabled runtime, enabling practical comparison of backend preparation, first-token behavior, speech handoff, and interaction semantics.

3) We build a deployment-oriented evaluation bench that measures backend runtime, media transport, terminal-level E2E latency, and interaction behavior for real-time video VLM applications, with public benchmark scores used only as model capability context.

## RELATED WORK

### Video Understanding And Streaming VLM Mechanisms

General video-language research has established strong capability in clip-level and long-video understanding. Surveys by Tang et al. (1) and Wu et al. (2) summarize progress in temporal reasoning, long-form video comprehension, and multimodal video-language modeling, while Video-LLaVA (Lin et al. (22)) and LLaVA-Video (Zhang et al. (23)) improve offline video QA and captioning through unified visual representation and video instruction tuning. These works are important foundations, but their dominant setting assumes that video evidence is available before the answer is produced.

Streaming video understanding changes the problem setting. A model must process visual evidence as it arrives, maintain bounded context, and answer before the interaction loses timeliness. This shift motivates not only better video understanding, but also timely state maintenance and deployable speech-and-media interaction. Streaming video VLMs differ in how they transform an unbounded media stream into model-ready context. Figure 2 abstracts the model-side design space into online video encoding, where incoming frames are converted into bounded visual state, and query- or event- triggered response, where text or event triggers are fused with current or retrieved visual context for decoding.

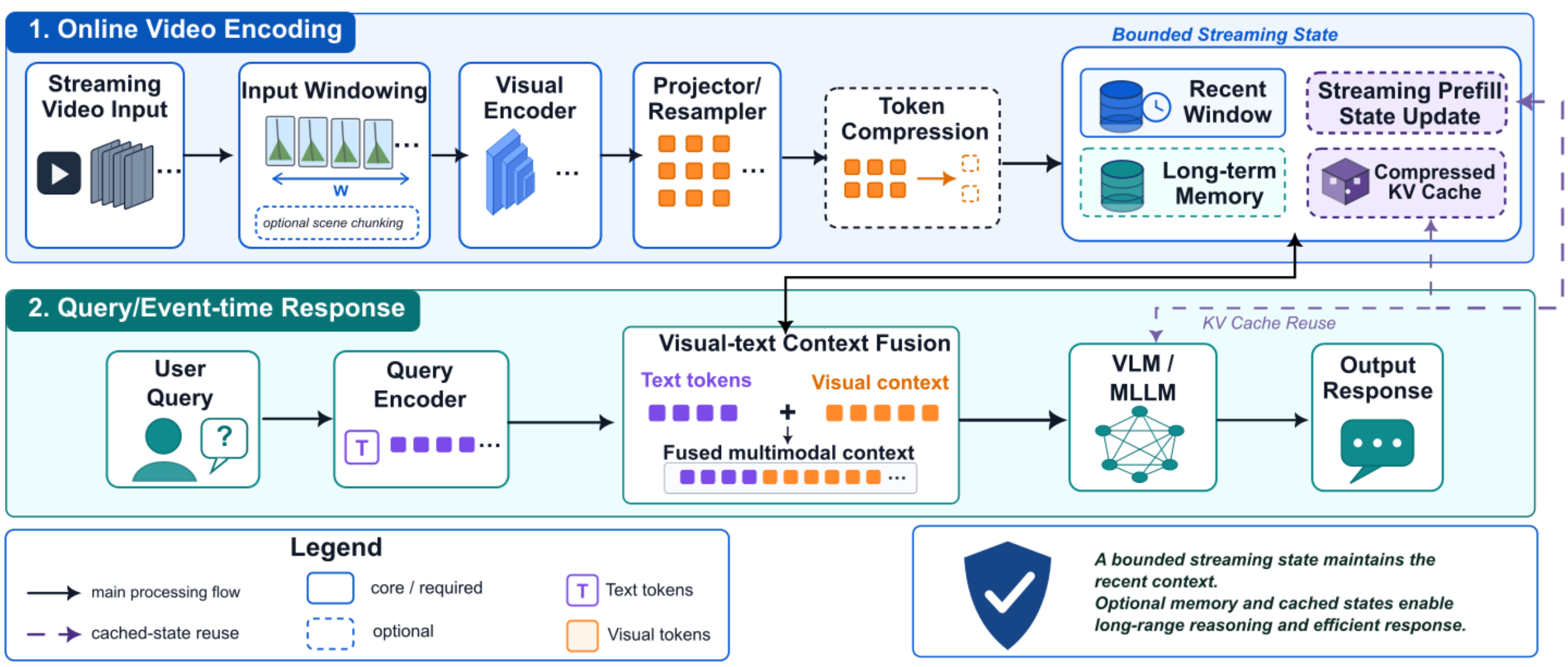

Figure 2 – General processing paradigm of streaming video VLMs. Solid modules denote the core path, and dashed modules denote architecture-dependent mechanisms such as scene chunking, token compression, streaming prefill, compressed KV cache, and cache reuse.

| Work | Scene/ window chunking | Token compression | Long-term memory | Streaming prefill/ state update | KV/ cache reuse | Event-triggered response |
|---|---|---|---|---|---|---|
| InfiniteVL (9) | × | × | √ | √ | × | × |
| ViSpeak (13) | √ | × | × | √ | √ | √ |
| MMDuet2 (14) | × | × | √ | √ | √ | √ |
| FluxMem (10) | √ | √ | √ | × | × | × |
| MiniCPM-o (12) | √ | √ | × | √ | √ | √ |
| AURA (6) | √ | × | √ | √ | √ | √ |

Table 1 – Mechanism matrix for representative streaming or interaction-oriented VLM studies.

Prior systems instantiate different parts of this flow. Some work targets active interaction and response timing, including Dispider (Qian et al. (15)), ViSpeak (Fu et al. (13)), MMDuet2 (Wang et al. (14)), AURA (Lu et al. (6)), and StreamReady (Azad et al. (16)). Other work focuses on efficient stream processing through token redundancy reduction, cache or retrieval support, sparse or linear attention, and memory compression, including TimeChat-Online (Yao et al. (5)), LiveVLM (Ning et al. (11)), InfiniteVL (Tao et al. (9)), StreamingVLM (Xu et al. (4)), FluxMem (Xie et al. (10)), and hierarchical token-compression methods (Wang et al. (25)). Streaming Video Instruction Tuning (Xia et al. (24)), Speak While Watching (Lin et al. (8)), and MiniCPM-o (OpenBMB (12)) further connect video streaming with real-time assistant behavior. Table 1 maps selected works to the optional modules in Figure 2.

### Streaming Evaluation Benchmarks

Benchmarks reflect the same shift from offline video understanding to online interaction. Video-MME (Fu et al. (3)) covers broad offline video analysis, while OVO-Bench (Niu et al. (20)) and StreamingBench (Lin et al. (21)) evaluate online and streaming comprehension. RIVER (Shi et al. (7)) frames real-time interaction around retrospective memory, live perception, and proactive response. We use these benchmarks as capability context and motivation for qualitative categories, not as reproduced accuracy scores.

### Real-Time Media Transport

Deployment requires real media transport and user-facing response paths. WebRTC provides browser-native, low-latency communication through integrated media transport, data channels, and NAT traversal. In contrast, RTSP manages session control without carrying media itself and is widely used in surveillance, video-on-demand, and professional media systems. For interactive VLMs, model-side streaming alone cannot ensure low latency, because endpointing, buffering, delivery, decoding, and playback also affect user-observed response. We therefore evaluate transport latency separately from backend inference and QA latency.

## SYSTEM ARCHITECTURE

### Edge-Cloud Service Overview

We design a server-centric edge-cloud system for low-latency speech-driven interaction over live video streams. Client devices remain lightweight by capturing audio/video, publishing media, displaying returned text, and playing synthesized speech. The cloud runtime maintains sessions, applies shared speech services, routes visual context and user queries to a selected VLM backend, and records measurement events. Figure 3 details the three-layer architecture: client surfaces, transport paths, and a unified service layer containing the VLM backend pool and archive-backed measurement. Phone, smart glasses, and PC clients represent deployment terminals, while the pseudo-replay mode supports repeatable backend and transport experiments. WebRTC, RTSP/WS, and pseudo paths are normalized into the same session representation before reaching the backend adapter interface.

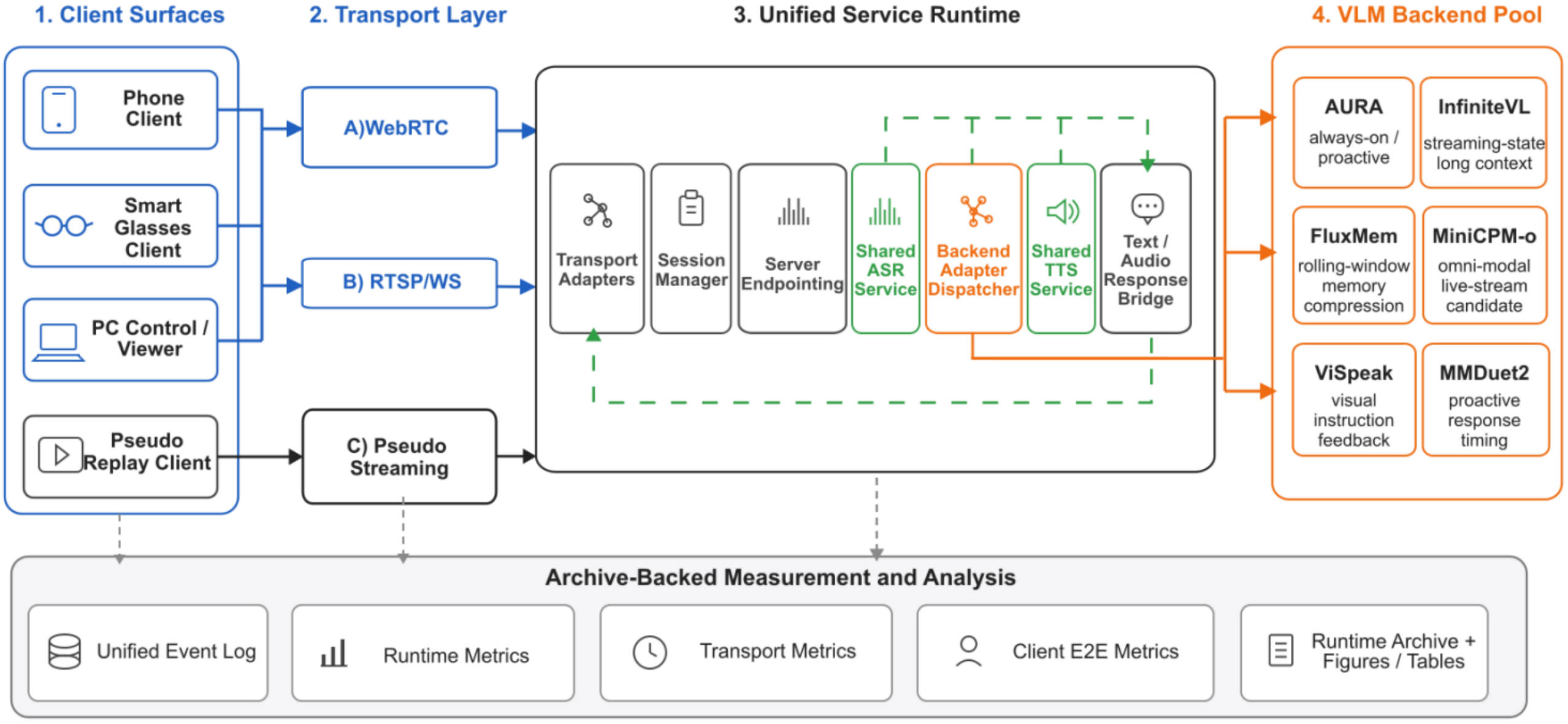

Figure 3 – Unified edge-cloud service architecture.

### Unified Runtime Flow

The runtime exposes a common speech-in, video-context, text/audio-out contract. Transport adapters normalize WebRTC, RTSP/WS, and pseudo replay into session events. The session manager tracks stream state, user turns, backend selection, and response delivery, endpointing identifies committed speech and establishes anchors for ASR and latency measurement. Shared ASR and TTS services sit outside the VLM backends. ASR converts committed speech into query text, the dispatcher sends the query and current visual context to the selected backend, and generated text is passed to shared TTS before the response bridge returns text and audio through the active transport path. This preserves one user-facing interaction contract while varying backend and transport choices.

### Adapting Heterogeneous VLM Backends

The backend pool includes AURA, InfiniteVL, FluxMem, MiniCPM-o, ViSpeak, and MMDuet2. Each receives shared-ASR query text and runtime-maintained visual context through a common adapter, while speech processing remains outside the backend. Integration must reconcile differences in visual-context organization, inference triggers, response timing, and text exposure. Backends may reuse streaming state, prepare rolling windows or compressed memory, materialize context at query time, or support background and event-driven processing.

Consequently, preparation, TTFT, VLM-to-TTS, and first-audio latency vary despite shared ASR/TTS and transport. InfiniteVL reuses streaming state, FluxMem performs query-time memory preparation, AURA and MMDuet2 support background or event-triggered responses, and MiniCPM-o and ViSpeak retain model-side streaming or interaction capabilities under the shared text-query interface.

### Archive-Backed Measurement

Measurement is built into the service path. The archive-backed module records unified runtime events and derives backend runtime, transport, and client-observed E2E metrics from the same runtime that serves interaction. This separation lets the deployment-oriented evaluation bench attribute latency to backend execution, transport delivery, or user-observed response behavior.

## IMPLEMENTATION

### Platform And Devices

The prototype runs VLM inference on one NVIDIA RTX 5880 Ada Generation GPU and shared ASR/TTS on another GPU of the same model, separating speech services from backend-specific VLM processes. Client surfaces include an Android phone, RayNeo X2 smart glasses, a PC control/viewer, and pseudo replay. The first three provide capture, display, and playback terminals, while pseudo replay provides repeatable server-side input.

### Runtime And Media Stack

The Python server runtime contains modules for media transport, backend invocation, shared speech services, and measurement instrumentation. WebRTC is implemented with LiveKit media tracks, data or transcription-style messages, and a long-lived server audio track for TTS. RTSP/WS uses FFmpeg and MediaMTX for RTSP audio/video publishing, receiving, and conversion, with WebSocket carrying text events because RTSP has no native ASR/VLM data channel. Shared ASR/TTS services use Qwen/Qwen3-ASR-1.7B and Qwen/Qwen3-TTS-12Hz-1.7B-Base, keeping backend comparisons independent of backend-specific speech stacks. Android clients use native capture and playback, while PC and pseudo clients support controlled protocol and replay experiments.

### Backend Resources And Wrappers

The six VLM backends are integrated from public releases, author-associated implementations, official model resources, or documented project materials. AURA, InfiniteVL, FluxMem, ViSpeak, and MMDuet2 use public research implementations, while MiniCPM-o uses the OpenBMB official model and documentation. The unified runtime connects upstream projects through service-layer wrappers. All wrappers receive shared-ASR text queries and runtime-prepared visual context; backend-native speech or duplex paths do not replace the shared ASR/TTS path.

### Runtime Modes And Measurement Records

The implementation exposes pseudo replay for backend profiling, RTSP/WS for media-style transport experiments, and WebRTC for full-duplex interaction. Each mode records structured runtime events and archive-backed summaries.

## EXPERIMENTS

### Evaluation Design And Metrics

The evaluation follows the system path from backend inference to user-observed interaction. Backend runtime decomposes the server-side speech-to-audio path into speech commitment to ASR final, ASR final to model-ready input, backend request to first VLM token, first usable model text to TTS submission, and TTS submission to first audio chunk. Transport measurements isolate one-way uplink and downlink latency for video, audio, and text. Client E2E measurements use client speech end as the anchor and report ASR final text, first VLM text, and first non-silent TTS audio. Model capability analysis uses author-reported streaming benchmark scores and qualitative same-video response examples.

### Backend Runtime Evaluation

Using pseudo transport, we isolate the service-internal speech-to-audio path from device and network effects while keeping orchestration, ASR/TTS services, scheduled queries, and generation limits fixed. As shown in Table 2, ASR-final and TTS-first latencies remain similar across backends, while preparation, VLM TTFT, and VLM-to-TTS handoff account for most differences. ViSpeak, InfiniteVL, AURA, and MiniCPM-o achieve first audio within 0.82–1.13 s, compared with 4.93 s for FluxMem and 17.39 s for MMDuet2.

The slower results primarily reflect backend adaptation costs. FluxMem incurs substantial query-time preparation and VLM inference latency, whereas MMDuet2 adds larger preparation, VLM TTFT, and output-handoff delays when adapted to immediate speech queries. In contrast, InfiniteVL minimizes preparation through reusable streaming state, although its VLM-to-TTS handoff remains relatively high.

### Transport Protocol Latency

The transport experiment isolates one-way application-observable latency for uplink and downlink video, audio, and text under WebRTC and RTSP/WS. Although the deployed interaction path mainly uses uplink video, downlink text, and bidirectional audio, measuring all directions gives a fuller protocol comparison. WebRTC denotes the LiveKit implementation, and RTSP/WS denotes RTSP media with WebSocket text events.

Timestamps are embedded in actual video frames, audio markers, and text messages. The analysis uses a 240 s official window, with NTP offsets below 0.12 ms and HTTP clock-correction p95/max offset of 1.5 ms.

| Backend | First audio (ms) | ASR (ms) | Prepare (ms) | VLM TTFT (ms) | VLM ->TTS (ms) | TTS First (ms) |
|---|---|---|---|---|---|---|
| AURA | 1057.4 | 85.1 | 474.1 | 146.2 | 130.6 | 201.6 |
| MiniCPM-o | 1125.2 | 85.4 | 442.4 | 45.1 | 338.9 | 213.5 |
| ViSpeak | 816.4 | 86.4 | 292.1 | 95.1 | 146 | 196.6 |
| InfiniteVL | 849.1 | 80.6 | 14.9 | 70.3 | 468.4 | 214.9 |
| FluxMem | 4925.9 | 85.9 | 2253.3 | 2154.6 | 234.4 | 197.7 |
| MMDuet2 | 17390.5 | 98.0 | 7471.9 | 2679.4 | 715.7 | 197.6 |

Table 2 – Backend runtime evaluation under pseudo-replay mode.

Figure 4 and Table 3 show a stable separation between the paths. WebRTC keeps video in the tens-of-milliseconds range, while RTSP/WS video remains near the half-second range; audio follows the same direction. Text remains low in both paths because RTSP/WS carries text over WebSocket. Near-zero or 0 ms text samples indicate delays below millisecond-level measurement and clock-correction precision, not physically zero transmission time.

| Transport | Direction | Track | Mean (ms) |
|---|---|---|---|
| WebRTC | Uplink | video | 39.429 |
| | | audio | 70.505 |
| | | text | 2.094 |
| | Downlink | video | 50.268 |
| | | audio | 62.828 |
| | | text | 2.879 |
| RTSP/WS | Uplink | video | 475.155 |
| | | audio | 263.665 |
| | | text | 0.162 |
| | Downlink | video | 505.882 |
| | | audio | 274.925 |
| | | text | 4.729 |

Table 3 – Transport latency summary

**Client-In-The-Loop Interaction Latency**

Table 4 summarizes a 15-turn client-in-the-loop evaluation across Android phone, smart glasses, and PC clients using MiniCPM-o and the same replayed speech-video QA task. With client speech end as the anchor, WebRTC provides similar performance across devices, achieving first-text latency of 0.92–0.98 s and first non-silent audio latency of 1.26–1.53 s. RTSP/WS increases these ranges to 1.18–1.19 s and 1.88–2.16 s, respectively. Because the server-side ASR, VLM, and TTS pipeline remains unchanged, the difference mainly reflects endpointing, buffering, media delivery, decoding, and playback.

**Streaming Capability Context And Interaction Examples**

Latency alone does not describe model capability or interaction quality. Table 5 reports author-provided OVO-Bench and StreamingBench scores where comparable coverage is available. MMDuet2 and InfiniteVL are excluded because their reports do not cover both benchmarks comparably; MMDuet2 reports only the StreamingBench CU proactive-output subtask, with a score of 34.69. These are cited results, not reproduced scores.

We then compare interaction behavior under the deployed service loop. Following RIVER (Shi et al. (7)), Figure 5 sorts matched same-video examples into real-time understanding, retrospective memory, proactive response, and multi-response behavior.

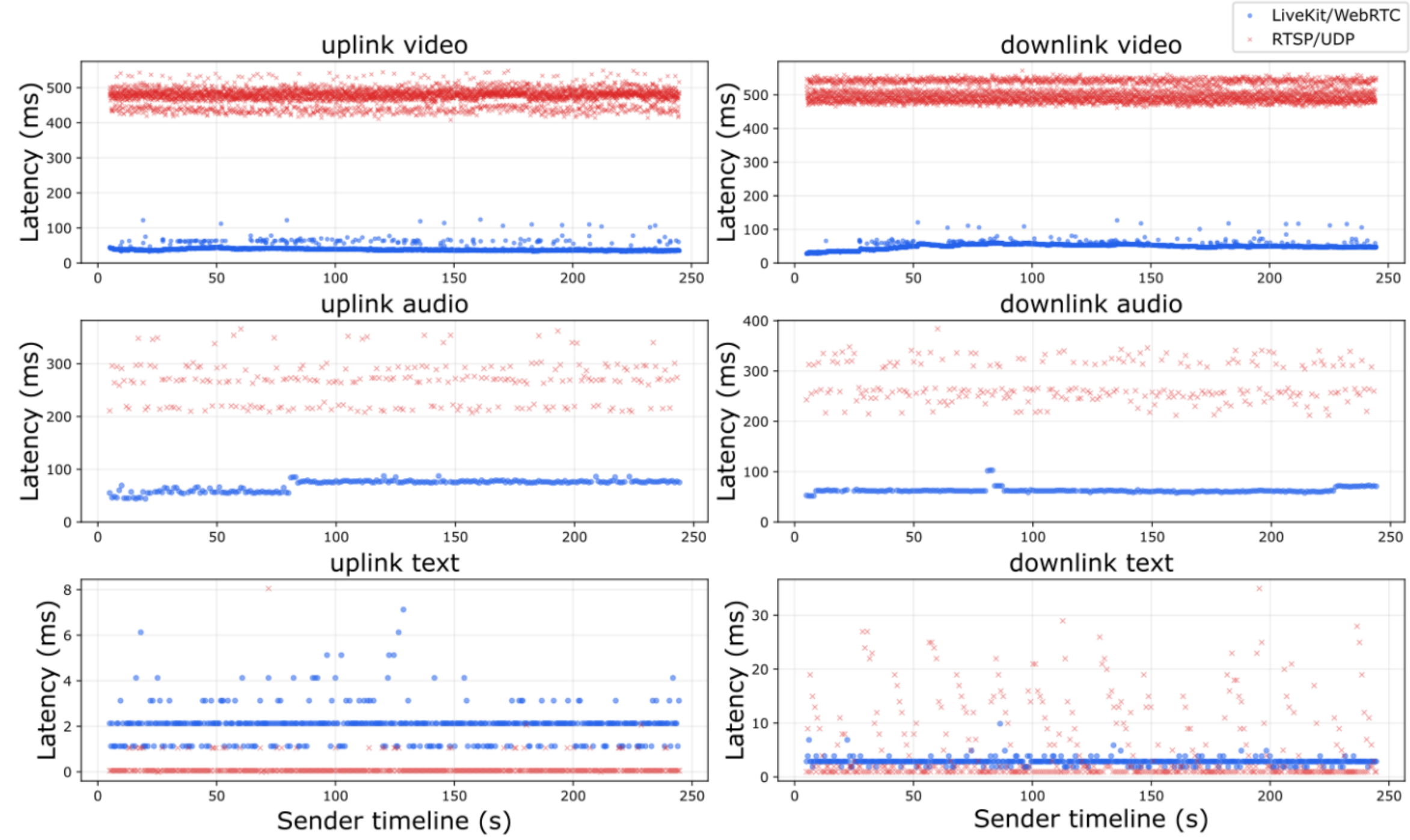

Figure 4 – Transport latency over the 4-minute official analysis window for LiveKit/WebRTC and RTSP/WS media with WebSocket text.

| Client | Transport | Turns | ASR final text(ms) | VLM first text(ms) | First non-silent TTS audio(ms) |
|---|---|---|---|---|---|
| Phone | WebRTC | 15 | 274.5 | 956.5 | 1272.9 |
| | RTSP/WS | | 567.7 | 1185.7 | 2049.9 |
| PC | WebRTC | | 272.0 | 919.5 | 1525.1 |
| | RTSP/WS | | 561.9 | 1177.9 | 1877.2 |
| Smart glasses | WebRTC | | 290.3 | 981.8 | 1258.4 |
| | RTSP/WS | | 554.8 | 1188.5 | 2157.9 |

| Method | OVO-Bench | | | | StreamingBench | | | |
|---|---|---|---|---|---|---|---|---|
| | RTVP | BT | FAR | ALL | RTVU | OSU | CU | ALL |
| Proprietary Models | | | | | | | | |
| GPT-4o | 64.5 | 60.8 | 53.4 | 59.5 | 73.3 | 44.5 | 38.7 | 60.2 |
| Gemini-1.5-Pro | 69.3 | 62.5 | 57.2 | 63.0 | 75.7 | 60.2 | 48.7 | 67.1 |
| Open-Source Models | | | | | | | | |
| FluxMem | 67.2 | - | - | - | 76.4 | - | - | - |
| Qwen3-VL-8B | 61.5 | 41.0 | 37.7 | 46.8 | 74.1 | 41.9 | 39.8 | 59.3 |
| ViSpeak | 66.3 | 57.5 | 54.3 | 61.1 | 70.4 | 61.6 | 43.9 | 62.6 |
| MiniCPM-o-4.5 | 67.3 | 55.9 | **55.9** | 59.7 | 78.2 | 42.1 | 44.5 | 62.7 |
| AURA | **79.8** | **60.4** | 55.8 | **65.3** | **83.2** | **62.0** | **59.0** | **73.1** |

Table 5 – Author-reported capability on OVO-Bench and StreamingBench. Boldface marks the best open-source score for each metric.

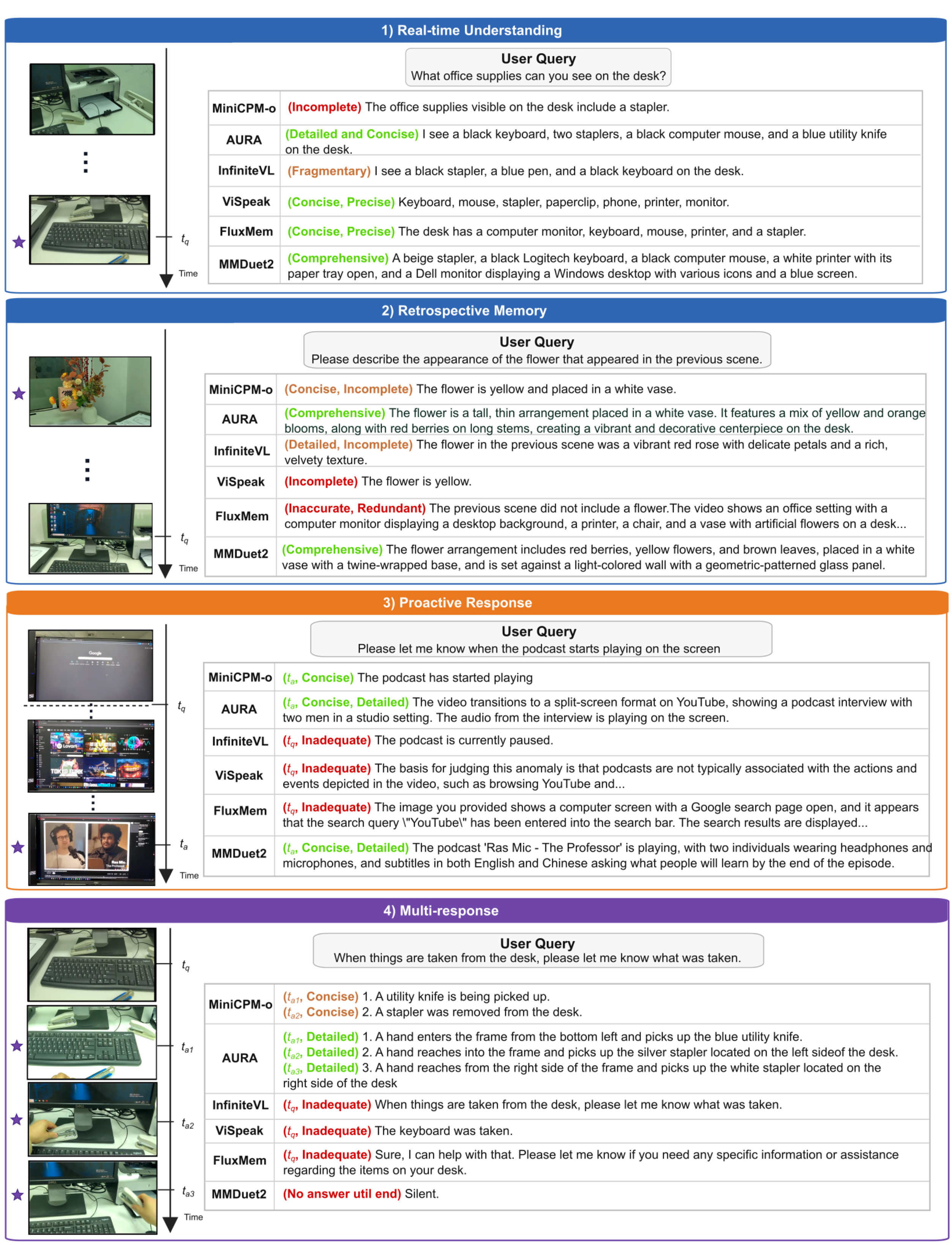

Figure 5 – Qualitative examples of six backend responses under matched video and prompt or trigger conditions. Real-time understanding asks about the current visual scene; retrospective memory asks about an earlier event; proactive response asks the system to respond when a future cue appears; multi-response asks the system to respond repeatedly as multiple relevant events occur.

Each model response in Figure 5 is preceded by compact labels that summarize the observed response quality. Green, orange, and red indicate good, moderate, and poor behavior, respectively. The notation $t_q$ marks query time, at $t_a$ marks a response near the relevant action or event, and $t_{a1}$, $t_{a2}$, and $t_{a3}$ denote the three removal events in the multi-response case. These labels support qualitative reading and are not formal numerical scores. The examples are illustrative rather than statistical. They show that current-scene understanding is broadly supported, whereas retrospective memory, proactive response, and multi-response interaction expose larger differences in temporal state retention and response scheduling. MMDuet2 is designed for interactive response timing, but its default 0.5 fps sampling can miss short action evidence; in this example it does not produce a valid multi-response output. Together, Table 5 and Figure 5 separate public benchmark context from deployed-service observations under a common speech-enabled runtime.

## DISCUSSION

The experiments show that real-time video VLM deployment is a layered system problem. Author-reported capability does not directly determine user-perceived quality because backend preparation, response exposure, transport, playback, and speech feedback can dominate different stages. For speech-feedback systems, VLM-to-TTS handoff is important because low TTFT alone does not ensure fast spoken feedback. Qualitative examples also show that interaction requires temporal evaluation beyond representative responses. Figure 5 illustrates retrospective, event-triggered, and multi-response behavior but does not measure alignment with key video evidence. Practical evaluation should cover response delay relative to action intervals, retrospective memory span, long-term trigger reliability, and completeness or duplication across repeated events.

This work has several limitations. OVO-Bench and StreamingBench scores are cited from original papers or official reports and are not reproduced in this framework. The qualitative labels are reading aids, not formal scores. Client-in-the-loop experiments use MiniCPM-o to control backend variation and focus on terminal and transport effects. Finally, the RTSP/WS path has not been deeply optimized for buffering, encoding, decoding, or player scheduling.

## CONCLUSION

We presented a low-latency edge-cloud system that integrates six VLM backends through shared ASR/TTS, transport, orchestration, adaptation, and logging interfaces. The system separately measures backend runtime, protocol latency, client-observed latency, and interaction behavior, achieving approximately 0.9–1.0 s to first VLM text and 1.3–1.5 s to first non-silent TTS audio feedback with suitable backend and transport choices. Future work will cover more backend-client combinations, diverse real-world scenes, optimized RTSP/WS delivery, and systematic metrics for retrospective, proactive, and multi-response interaction.

## ACKNOWLEDGEMENTS

This work was partly supported by the NSFC (62431015, 62571317, 62501387), the Fundamental Research Funds for the Central Universities, Shanghai Key Laboratory of Digital Media Processing and Transmission under Grant 22DZ2229005, Special Fund for Promoting High-Quality Industrial Development (2025358) and the 111 Project BP0719010.